\documentclass[sigconf]{acmart}

\AtBeginDocument{%
  }

\copyrightyear{2024}
\acmYear{2024}
\setcopyright{acmlicensed}\acmConference[MOBILESoft '24]{IEEE/ACM 11th International Conference on Mobile Software Engineering and Systems}{April 14--15, 2024}{Lisbon, Portugal}
\acmBooktitle{IEEE/ACM 11th International Conference on Mobile Software Engineering and Systems (MOBILESoft '24), April 14--15, 2024, Lisbon, Portugal}
\acmDOI{10.1145/3647632.3651392}
\acmISBN{979-8-4007-0594-6/24/04}

\usepackage{setspace}
\usepackage{algorithmic}
\usepackage{graphicx}
\usepackage{textcomp}
\usepackage{xcolor}
\usepackage{soul}
\usepackage{comment}
\usepackage[linesnumbered,ruled,vlined]{algorithm2e}
\usepackage{balance}
\usepackage{microtype}
\usepackage{multirow}
\usepackage[normalem]{ulem}
\usepackage{soul}
\usepackage{lipsum}
\usepackage{wrapfig}
\usepackage{graphicx}
\usepackage{tikz}
\usepackage{dblfloatfix} 

\newcommand*\circled[1]{\tikz[baseline=(char.base)]{\small{\textbf{
			\node[shape=circle,fill,inner sep=0.75pt] (char) {\textcolor{white}{#1}};}}}}

\definecolor{amethyst}{rgb}{0.6, 0.4, 0.8}

\newcommand{\toolname}{\textbf{PauseNow}}

\begin{document}

\title{Digital Wellbeing Redefined: Toward User-Centric Approach for Positive Social Media Engagement}

\author{Yixue Zhao}
\email{yzhao@isi.edu}
\orcid{0000-0003-3046-6621}
\affiliation{%
  \institution{USC Information Sciences Institute}
  \city{Arlington}
  \state{Virginia}
  \country{USA}
}

\author{Tianyi Li}
\email{li4251@purdue.edu}
\orcid{0000-0002-1145-2526}
\affiliation{%
  \institution{Purdue University}
  \city{West Lafayette, Indiana}
  \country{USA}}

\author{Michael Sobolev}
\email{michael.sobolev@cshs.org}
\orcid{0000-0002-8931-7682}
\affiliation{%
  \institution{University of Southern California}
  \city{Los Angeles, California}
  \country{USA}}

\begin{abstract}
The prevalence of social media and its escalating impact on mental health has highlighted the need for effective digital wellbeing strategies. 
Current digital wellbeing interventions have primarily focused on reducing screen time and social media use, often neglecting the potential benefits of these platforms. 
This paper introduces a new perspective centered around empowering positive social media experiences, instead of limiting users with restrictive rules.
In line with this perspective, we lay out the key requirements that should be considered in future work, aiming to spark a dialogue in this emerging area.
We further present our initial effort to address these requirements with \toolname, an innovative digital wellbeing intervention designed to align users' digital \emph{behaviors} with their \emph{intentions}. \toolname~ leverages digital nudging and intention-aware recommendations to gently guide users back to their original intentions when they ``get lost'' during their digital usage,  promoting a more mindful use of social media. 
\end{abstract}


\begin{teaserfigure}
  \includegraphics[width=\textwidth]{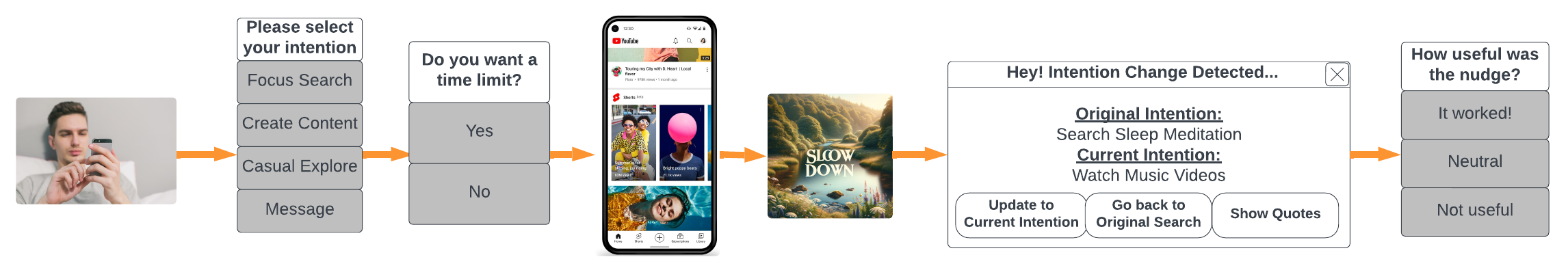}
    \caption{An example use case of how a user may interact with \toolname. Note that \toolname~ is co-designed with the user, thus the user interfaces (e.g., how to nudge the user, options presented when intention drifts) may vary across different users.}
	\label{fig:use_case}
\end{teaserfigure}

\maketitle


\section{Introduction}
\label{sec:intro}

In the digital era, the prevalence of social media use, particularly among youth, has raised critical concerns
about mental health and digital wellbeing~\cite{buchi2021digital, peper2018digitaladdiction, allcott2022digital, thomas2022digital}. As we navigate through an increasingly connected world, the impact
of digital usage on our psychological and emotional states becomes a crucial area of research~\cite{rad2019youth, buchi2021digital, cecchinato2019designing, monge2019race, burr2020ethics, roffarello2023achieving}. Current
approaches to digital wellbeing have primarily focused on reducing social media use by incorporating
friction, limiting screen time, and altering app aesthetics to make it less engaging~\cite{orzikulova2023finerme, lyngs2020justhack, roffarello2023achieving, zimmermann2023digital}. However, these methods often overlook the
multifaceted nature of social media---a platform that, when used \emph{mindfully}, can offer significant benefits such as
fostering connections, staying informed, and even enhancing mental health via uplifting content~\cite{buchi2021digital}.
This paradoxical nature of social media necessitates a more nuanced approach to digital wellbeing.
Limiting use without considering the context may not only be ineffective but could also provoke counterproductive
behaviors, as users often seek a sense of agency and control over their digital experiences~\cite{lukoff2021design}. 
Inspired by positive computing research~\cite{calvo2014positive}, we propose a paradigm shift towards empowering users for \emph{positive} social media engagement. 
This concept of positive engagement encompasses three levels: (1) \emph{self-awareness}, which involves understanding one's social media usage and aspirations for improvement; 
(2) \emph{alignment} of personal goals with actual digital behaviors; 
and (3) enhancement of \emph{long-term wellbeing}.
This perspective moves away from punitive measures for unintended behaviors, such as restricting app usage, which can frequently result in feelings of guilt, regret, and dissatisfaction~\cite{cho2021reflect}.

To capture this perspective, we redefine digital wellbeing as \emph{a harmonious alignment between a user's intentions and their digital behaviors to enhance long-term wellbeing}. 
In line with our definition, we design \underline{\toolname}, a \underline{\textbf{P}}ersonalized, \underline{\textbf{a}}daptive, \underline{\textbf{use}}r-centric i\underline{\textbf{N}}tervention f\underline{\textbf{o}}r
digital \textbf{w}ell-being. \toolname~ discerns discrepancies between intended and actual digital usage and offers
customized support based on individual preferences. Utilizing the digital nudging and intention-aware recommendations ~\cite{okeke2018towards, sobolev2021digitalnudging, yang2019intention,chen2019air, knijnenburg2016recommender},
\toolname~ reminds users of their original intentions when they veer off course, providing them with full autonomy to tailor the intervention to their specific needs. 
For instance, if \toolname~ detects a user's shift from a focused information search to aimless browsing, a user-defined system behavior will be triggered to help the user re-evaluate their intentions in the app. These user-defined system behaviors are customizable, such as a prompt recalling the original search query, displaying the current time, or showing the user's daily goals. It aims to foster mindful interactions with digital platforms, countering social media app's recommendation algorithms that typically aim to maximize screen time.

As the issue of digital wellbeing grows increasingly vital and widespread~\cite{flipphone, monge2019race, roffarello2023achieving}, its exploration within the Mobile Software Engineering community remains notably limited. 
Recent efforts~\cite{10.1145/3565066.3608703,10.1145/3582515.3609523}  have begun to explore this new perspective to support conscious social media use through nudging technology, demonstrating initial promise. 
However, these interventions are currently preliminary and limited in scope, focusing on specific scenarios such as ``infinite scroll'' and ``pull to refresh'' in the newsfeeds~\cite{10.1145/3565066.3608703}. 
This gap has motivated us to outline important considerations in this emerging field, as well as design \toolname~ as a reference framework with broader applicability to guide future research.
We aim to spark a dialogue on this important topic, harnessing our collective knowledge and expertise in this research forum.

\section{Intervention Requirements}
\label{sec:req}

This section outlines five key requirements for designing effective digital wellbeing interventions.

\begin{figure*}[!t]
	\includegraphics[width=\textwidth]{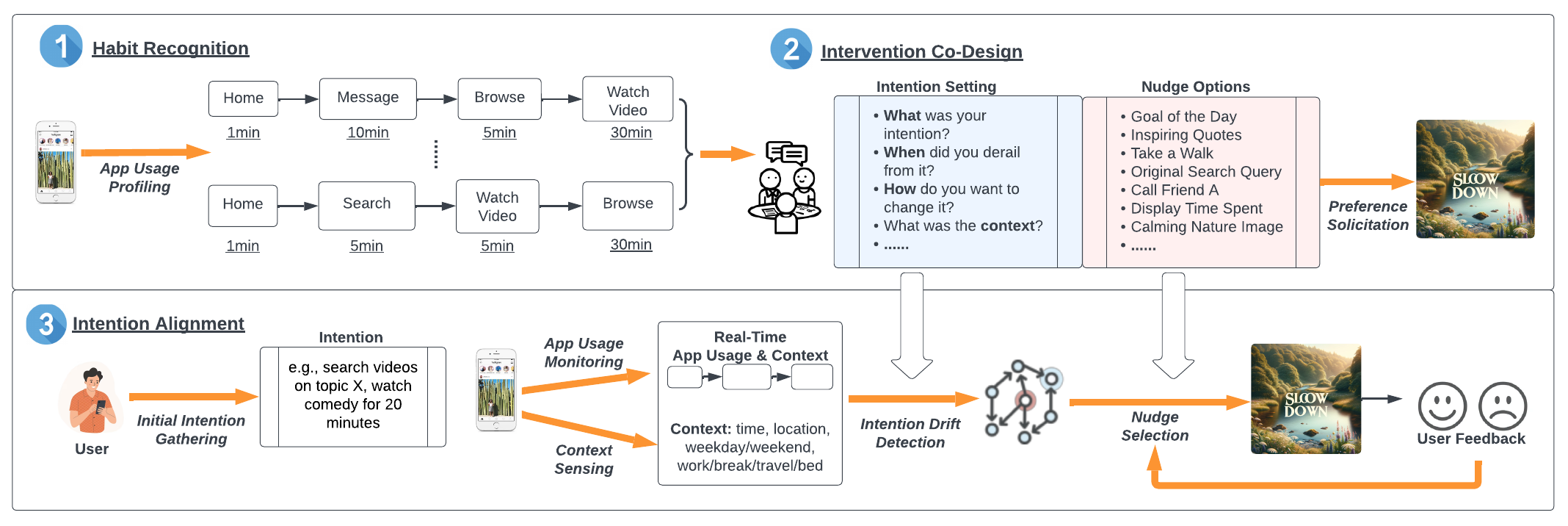}
    \caption{Overview of \toolname's three-phase workflow. (1) {Habit Recognition} detects habitual patterns to bring digital awareness to the user. (2) {Intervention Co-Design} solicits user preferences to customize \toolname's design. (3) Intention Alignment detects intention drifts and brings the user back to their intended usage using their preferred method.}
	\label{fig:design}
\end{figure*}

\textbf{Req$_1$ --- \emph{The intervention design shall be context-aware and user-centric to reflect both inter- and intra-individual differences. --- }}
\noindent 
Research has shown considerable \emph{inter-individual} variability in digital behaviors, as users interact with apps in distinct ways based on their unique purposes and habits~\cite{lukoff2018meaningful, buchi2021digital}. 
For example, while one user might struggle with online forums such as \texttt{Reddit}, another could find video platforms such as \texttt{YouTube} challenging to manage. 
Additionally, variations in individual personality and preferences call for {personalized} solutions as users often find different digital wellbeing tools effective~\cite{lyngs2022goldilocks, lyngs2019self}. 
Furthermore, there are significant \emph{intra-individual} variations in users' digital behaviors largely influenced by context, with the same user's intentions frequently changing over time, even within the same app~\cite{lukoff2021design, lukoff2023switchtube}. 
For example, a user might engage with a video platform for educational content during working hours, yet later uses the same platform for leisure purposes when taking a break, such as watching comedy videos.  
Inaccurate interpretation of the user's context can lead to misalignment between the intervention and the user's actual intentions, potentially causing frustration and reduced engagement with the intervention. 
Thus, it is essential for digital wellbeing tools to consider \emph{context-awareness} and \emph{personalization} to continuously adapt to the user's evolving circumstances and behavior patterns.

\textbf{Req$_2$ --- \emph{The intervention shall foster self-awareness and promote healthy digital habits. --- }}
\noindent 
A key challenge in digital wellbeing is not merely the user's inability to make informed decisions, but rather the lack of \emph{awareness} during digital use, which often leads to auto-pilot behaviors such as \emph{mindlessly} browsing content driven by platform's recommendation algorithms. 
To enhance digital wellbeing in the long term, interventions must improve the user's self-awareness in digital use and also assist them in cultivating healthier digital habits. Digital nudging that we incorporate in \toolname~ has shown usefulness in fostering positive digital behavior habits~\cite{sobolev2021digitalnudging, okeke2018goodvibes, okeke2018towards, caraban201923}.
Additionally, considering the significant impact of environmental factors on habit formation, effective interventions must be flexible and adapt to changing contexts (e.g., work vs. vacation) to maintain their relevance and efficacy.

\textbf{Req$_3$ --- \emph{The intervention shall offer motivating alternatives that go beyond digital devices. --- }}
\noindent 
To enhance long-term wellbeing, the intervention shall extend beyond the confines of the digital realm and offer meaningful alternatives to digital engagement, such as encouraging real-world social interactions outside of digital devices. 
Aligning with Req$_1$ and Req$_2$, the alternatives should be tailored to the user's personal preferences (e.g., favorite social activities) and relevant context (e.g., social circle, time of day). 
This will enable the suggestion of high-quality, appealing alternatives that genuinely motivate users to disengage from their digital devices when beneficial for long-term wellbeing. 

\textbf{Req$_4$ --- \emph{The intervention shall strike a balance between user control and machine autonomy. --- }}
\noindent 
Users often desire control over their digital experiences, valuing a sense of agency~\cite{lukoff2021design, lukoff2023switchtube}. 
Yet, they can feel overwhelmed by the sheer volume of features and options typical digital wellbeing apps offer. Thus, the intervention's frequency, the variety and relevance of the options presented, and the contextual applicability must all be carefully calibrated. For instance, the digital wellbeing tool SwitchTube offers ``focus mode'' and ``explore mode'', providing varying degrees of user agency, and has shown user preferences for these modes change according to their context~\cite{lukoff2023switchtube}. Understanding the user's context is crucial to offering the proper granularity of control tailored to user needs.

\textbf{Req$_5$ --- \emph{The intervention shall mitigate the habituation effect. --- }}
\noindent 
Habituation can significantly reduce the effectiveness of interventions, leading users to routinely overlook or dismiss prompts, such as habitually ignoring a frequently encountered reminder~\cite{gotzian2023modeling, anderson2016warning, 10.1145/3582515.3609523}. Additionally, it may result in users deviating from their original intentions and mindlessly navigating to specific apps or app features out of habit.
To mitigate this, interventions must be designed with variability and adaptability, ensuring that reminders remain fresh and engaging. 
This might include varying the format, content, or timing of prompts. 
Moreover, incorporating a design that is both context-aware and intention-aware can offer sustained value to users, thereby fostering their long-term engagement with the tool.
Addressing the habituation effect is crucial not only for immediate effectiveness but also for maintaining the long-term efficacy and impact of the intervention.

\section{\toolname~ Approach}
\label{sec:appr}

Guided by the requirements above, we now introduce \toolname, a {user-centric} digital wellbeing intervention for enhancing positive social media engagement. Note that \toolname's current design is conceptual, serving as a reference framework for digital wellbeing interventions targeting a specific problem domain, such as video social media app \texttt{YouTube} or forum-based app \texttt{Reddit}. 
Figure~\ref{fig:design} depicts \toolname's three-phase workflow that we detail below.

\circled{1} \textbf{Habit Recognition} phase aims to recognize a user's digital habits and bring self-awareness, as the lack of awareness is one of the key obstacles in managing digital behaviors. 
This phase is integral in helping users gain insights and clarity into their digital behavior patterns versus their intended usage, setting the groundwork for the co-design sessions in the next phase.
Specifically, the Habit Recognition phase utilizes App Usage Profiling technique to extract a model of the user’s behavior within the app as illustrated in Figure~\ref{fig:design}. The behavior model is a finite state machine that represents the user's common behavior chains when using a particular app. 
The underlying techniques for constructing this model may vary, often involving a trade-off between accuracy and generalization. For example, FinerMe~\cite{orzikulova2023finerme} leverages the accessibility features of Android systems to extract UI hierarchy data for analyzing user behavior within specific applications, such as \texttt{YouTube} and \texttt{Instagram}. While this approach is highly accurate, it is limited in scope, applicable only to certain apps and the Android platform.
In contrast, our prior work \textsc{Avgust}~\cite{zhao2022avgust}, employs a more versatile, platform-independent approach using Computer Vision techniques to discern screen and widget semantics across a diverse range of applications. 
While this approach ensures broad applicability across different apps and operating systems,  the generated behavior model may be less accurate in some scenarios.

\circled{2} \textbf{Intervention Co-Design} phase conducts interviews with the user to understand their needs and solicits their requirements and preferences to guide \toolname's development. 
Note that while interviews form the basis of our initial study for deeper insights, we envision \toolname's co-design process to be automated without interviews in future iteration, such as offering preference settings and gathering user feedback to refine the design.
Insights from the Habit Recognition phase, especially the behavior patterns that lead to the user's intention drifts, are shared to enhance their self-awareness. 
Sample interview questions are illustrated in Figure~\ref{fig:design}. Responses obtained from these interviews are utilized to flag certain paths in the behavior model extracted from phase 1 as \emph{unintentional}.
These patterns are then used to aid the Intention Drift Detection module in the subsequent phase.
Furthermore, a range of Nudge Options, such as ``Goal of the Day'', ``Call Friend A'', or ``Calming Nature Imagery'' (as depicted in Figure~\ref{fig:design}), is presented to the user through various app prototypes for them to choose how they wish to be nudged when intention drift occurs. 
This approach is directly motivated by our redefinition of digital wellbeing that aims to align users' digital behaviors with their true intentions, promoting healthier digital habits rather than imposing out-of-context punitive measures such as forced app closures.
The chosen nudge options are then incorporated into the Nudge Selection module in the subsequent phase, allowing for the customization of the user interfaces in \toolname~ to suit individual preferences and needs.

\circled{3} \textbf{Intention Alignment} phase utilizes the artifacts obtained from the previous phases as mentioned earlier and prompts the user with their preferred nudge options when unintentional behaviors are detected. 
As Figure~\ref{fig:design} shows, \toolname~ first gathers the user's initial intention via user input at the launch of a social media app. 
Following app initiation, \toolname~ performs App Usage Monitoring and Context Sensing to monitor the real-time app usage and context, including what app feature the user interacts with and its associated context, such as \emph{<``Watch Video'', ``2PM'', ``Bed'', ``Weekday''>}. 
This process reuses App Usage Profiling from the Habit Recognition phase and augments the behavior model with real-time contextual information.
The Intention Drift Detection module then evaluates the behavior model to identify deviations from the user's initial intention, leveraging insights from the Intervention Co-Design phase.
When intention drift is detected, the Nudge Selection module intervenes, nudging the user based on their preferences solicited from the previous phase to steer them back towards their original intentions and foster mindful social media engagement.
To counter the ``habituation effect'', the Nudge Selection module periodically adjusts its nudges, learning from similar user groups and incorporating real-time user feedback for dynamic enhancement.
Figure~\ref{fig:use_case} presents a typical scenario depicting user interaction with \toolname~ during the Intention Alignment phase, demonstrating the system's practical application.
\section{Conclusion and Future Directions}
\label{sec:discussion}

In this paper, we redefine digital wellbeing, emphasizing positive and mindful social media engagement, and identify essential requirements aligned with this perspective.
We further introduce \toolname, a novel, user-centric approach for digital wellbeing that serves as a reference framework to guide future research. 
This work paves the way for a new class of digital wellbeing interventions, opening up a wide array of future opportunities, such as 
developing cross-platform techniques with broader applicability, 
balancing the trade-offs between user privacy and personalization, 
automating user preference elicitation for enhanced long-term engagement, and
developing multimodal AI algorithms to better understand the user's change of context. 
Furthermore, it is essential to continuously assess the impact of social media platforms on end users' digital wellbeing as their design can largely influence user behaviors.
These multifaceted challenges invite interdisciplinary expertise and collaboration to create a digital environment that not only respects user preferences but also promotes a healthier, more balanced relationship with technology.







\bibliographystyle{ACM-Reference-Format}
\bibliography{reference}


\begin{thebibliography}{32}


\ifx \showCODEN    \undefined \def \showCODEN     #1{\unskip}     \fi
\ifx \showDOI      \undefined \def \showDOI       #1{#1}\fi
\ifx \showISBNx    \undefined \def \showISBNx     #1{\unskip}     \fi
\ifx \showISBNxiii \undefined \def \showISBNxiii  #1{\unskip}     \fi
\ifx \showISSN     \undefined \def \showISSN      #1{\unskip}     \fi
\ifx \showLCCN     \undefined \def \showLCCN      #1{\unskip}     \fi
\ifx \shownote     \undefined \def \shownote      #1{#1}          \fi
\ifx \showarticletitle \undefined \def \showarticletitle #1{#1}   \fi
\ifx \showURL      \undefined \def \showURL       {\relax}        \fi
\providecommand\bibfield[2]{#2}
\providecommand\bibinfo[2]{#2}
\providecommand\natexlab[1]{#1}
\providecommand\showeprint[2][]{arXiv:#2}

\bibitem[Allcott et~al\mbox{.}(2022)]%
        {allcott2022digital}
\bibfield{author}{\bibinfo{person}{Hunt Allcott}, \bibinfo{person}{Matthew
  Gentzkow}, {and} \bibinfo{person}{Lena Song}.}
  \bibinfo{year}{2022}\natexlab{}.
\newblock \showarticletitle{Digital addiction}.
\newblock \bibinfo{journal}{\emph{American Economic Review}}
  \bibinfo{volume}{112}, \bibinfo{number}{7} (\bibinfo{year}{2022}),
  \bibinfo{pages}{2424--2463}.
\newblock


\bibitem[Anderson et~al\mbox{.}(2016)]%
        {anderson2016warning}
\bibfield{author}{\bibinfo{person}{Bonnie~Brinton Anderson},
  \bibinfo{person}{Anthony Vance}, \bibinfo{person}{C~Brock Kirwan},
  \bibinfo{person}{Jeffrey~L Jenkins}, {and} \bibinfo{person}{David Eargle}.}
  \bibinfo{year}{2016}\natexlab{}.
\newblock \showarticletitle{From warning to wallpaper: Why the brain habituates
  to security warnings and what can be done about it}.
\newblock \bibinfo{journal}{\emph{Journal of Management Information Systems}}
  \bibinfo{volume}{33}, \bibinfo{number}{3} (\bibinfo{year}{2016}),
  \bibinfo{pages}{713--743}.
\newblock


\bibitem[B{\"u}chi(2021)]%
        {buchi2021digital}
\bibfield{author}{\bibinfo{person}{Moritz B{\"u}chi}.}
  \bibinfo{year}{2021}\natexlab{}.
\newblock \showarticletitle{Digital well-being theory and research}.
\newblock \bibinfo{journal}{\emph{New Media \& Society}}
  (\bibinfo{year}{2021}), \bibinfo{pages}{14614448211056851}.
\newblock


\bibitem[Burr et~al\mbox{.}(2020)]%
        {burr2020ethics}
\bibfield{author}{\bibinfo{person}{Christopher Burr},
  \bibinfo{person}{Mariarosaria Taddeo}, {and} \bibinfo{person}{Luciano
  Floridi}.} \bibinfo{year}{2020}\natexlab{}.
\newblock \showarticletitle{The ethics of digital well-being: A thematic
  review}.
\newblock \bibinfo{journal}{\emph{Science and engineering ethics}}
  \bibinfo{volume}{26}, \bibinfo{number}{4} (\bibinfo{year}{2020}),
  \bibinfo{pages}{2313--2343}.
\newblock


\bibitem[Calvo and Peters(2014)]%
        {calvo2014positive}
\bibfield{author}{\bibinfo{person}{Rafael~A Calvo} {and}
  \bibinfo{person}{Dorian Peters}.} \bibinfo{year}{2014}\natexlab{}.
\newblock \bibinfo{booktitle}{\emph{Positive computing: technology for
  wellbeing and human potential}}.
\newblock \bibinfo{publisher}{MIT press}.
\newblock


\bibitem[Caraban et~al\mbox{.}(2019)]%
        {caraban201923}
\bibfield{author}{\bibinfo{person}{Ana Caraban}, \bibinfo{person}{Evangelos
  Karapanos}, \bibinfo{person}{Daniel Gon{\c{c}}alves}, {and}
  \bibinfo{person}{Pedro Campos}.} \bibinfo{year}{2019}\natexlab{}.
\newblock \showarticletitle{23 ways to nudge: A review of technology-mediated
  nudging in human-computer interaction}. In
  \bibinfo{booktitle}{\emph{Proceedings of the 2019 CHI conference on human
  factors in computing systems}}. \bibinfo{pages}{1--15}.
\newblock


\bibitem[Cecchinato et~al\mbox{.}(2019)]%
        {cecchinato2019designing}
\bibfield{author}{\bibinfo{person}{Marta~E Cecchinato}, \bibinfo{person}{John
  Rooksby}, \bibinfo{person}{Alexis Hiniker}, \bibinfo{person}{Sean Munson},
  \bibinfo{person}{Kai Lukoff}, \bibinfo{person}{Luigina Ciolfi},
  \bibinfo{person}{Anja Thieme}, {and} \bibinfo{person}{Daniel Harrison}.}
  \bibinfo{year}{2019}\natexlab{}.
\newblock \showarticletitle{Designing for digital wellbeing: A research \&
  practice agenda}. In \bibinfo{booktitle}{\emph{Extended abstracts of the 2019
  CHI conference on human factors in computing systems}}.
  \bibinfo{pages}{1--8}.
\newblock


\bibitem[Chen et~al\mbox{.}(2019)]%
        {chen2019air}
\bibfield{author}{\bibinfo{person}{Tong Chen}, \bibinfo{person}{Hongzhi Yin},
  \bibinfo{person}{Hongxu Chen}, \bibinfo{person}{Rui Yan},
  \bibinfo{person}{Quoc Viet~Hung Nguyen}, {and} \bibinfo{person}{Xue Li}.}
  \bibinfo{year}{2019}\natexlab{}.
\newblock \showarticletitle{Air: Attentional intention-aware recommender
  systems}. In \bibinfo{booktitle}{\emph{2019 IEEE 35th International
  Conference on Data Engineering (ICDE)}}. IEEE, \bibinfo{pages}{304--315}.
\newblock


\bibitem[Cho et~al\mbox{.}(2021)]%
        {cho2021reflect}
\bibfield{author}{\bibinfo{person}{Hyunsung Cho}, \bibinfo{person}{DaEun Choi},
  \bibinfo{person}{Donghwi Kim}, \bibinfo{person}{Wan~Ju Kang},
  \bibinfo{person}{Eun~Kyoung Choe}, {and} \bibinfo{person}{Sung-Ju Lee}.}
  \bibinfo{year}{2021}\natexlab{}.
\newblock \showarticletitle{Reflect, not regret: Understanding regretful
  smartphone use with app feature-level analysis}.
\newblock \bibinfo{journal}{\emph{Proceedings of the ACM on human-computer
  interaction}} \bibinfo{volume}{5}, \bibinfo{number}{CSCW2}
  (\bibinfo{year}{2021}), \bibinfo{pages}{1--36}.
\newblock


\bibitem[Gotzian(2023)]%
        {gotzian2023modeling}
\bibfield{author}{\bibinfo{person}{Lisa Gotzian}.}
  \bibinfo{year}{2023}\natexlab{}.
\newblock \showarticletitle{Modeling the decreasing intervention effect in
  digital health: a computational model to predict the response for a walking
  intervention}.
\newblock  (\bibinfo{year}{2023}).
\newblock


\bibitem[Hill(2024)]%
        {flipphone}
\bibfield{author}{\bibinfo{person}{Kashmir Hill}.}
  \bibinfo{year}{2024}\natexlab{}.
\newblock \bibinfo{title}{I Was Addicted to My Smartphone, So I Switched to a
  Flip Phone for a Month}.
\newblock
\newblock
\urldef\tempurl%
\url{https://www.nytimes.com/2024/01/06/technology/smartphone-addiction-flip-phone.html}
\showURL{%
\tempurl}


\bibitem[Knijnenburg et~al\mbox{.}(2016)]%
        {knijnenburg2016recommender}
\bibfield{author}{\bibinfo{person}{Bart~P Knijnenburg},
  \bibinfo{person}{Saadhika Sivakumar}, {and} \bibinfo{person}{Daricia
  Wilkinson}.} \bibinfo{year}{2016}\natexlab{}.
\newblock \showarticletitle{Recommender systems for self-actualization}. In
  \bibinfo{booktitle}{\emph{Proceedings of the 10th acm conference on
  recommender systems}}. \bibinfo{pages}{11--14}.
\newblock


\bibitem[Lukoff et~al\mbox{.}(2023)]%
        {lukoff2023switchtube}
\bibfield{author}{\bibinfo{person}{Kai Lukoff}, \bibinfo{person}{Ulrik Lyngs},
  \bibinfo{person}{Karina Shirokova}, \bibinfo{person}{Raveena Rao},
  \bibinfo{person}{Larry Tian}, \bibinfo{person}{Himanshu Zade},
  \bibinfo{person}{Sean~A Munson}, {and} \bibinfo{person}{Alexis Hiniker}.}
  \bibinfo{year}{2023}\natexlab{}.
\newblock \showarticletitle{SwitchTube: A Proof-of-Concept System Introducing
  “Adaptable Commitment Interfaces” as a Tool for Digital Wellbeing}. In
  \bibinfo{booktitle}{\emph{Proceedings of the 2023 CHI Conference on Human
  Factors in Computing Systems}}. \bibinfo{pages}{1--22}.
\newblock


\bibitem[Lukoff et~al\mbox{.}(2021)]%
        {lukoff2021design}
\bibfield{author}{\bibinfo{person}{Kai Lukoff}, \bibinfo{person}{Ulrik Lyngs},
  \bibinfo{person}{Himanshu Zade}, \bibinfo{person}{J~Vera Liao},
  \bibinfo{person}{James Choi}, \bibinfo{person}{Kaiyue Fan},
  \bibinfo{person}{Sean~A Munson}, {and} \bibinfo{person}{Alexis Hiniker}.}
  \bibinfo{year}{2021}\natexlab{}.
\newblock \showarticletitle{How the design of youtube influences user sense of
  agency}. In \bibinfo{booktitle}{\emph{Proceedings of the 2021 CHI Conference
  on Human Factors in Computing Systems}}. \bibinfo{pages}{1--17}.
\newblock


\bibitem[Lukoff et~al\mbox{.}(2018)]%
        {lukoff2018meaningful}
\bibfield{author}{\bibinfo{person}{Kai Lukoff}, \bibinfo{person}{Cissy Yu},
  \bibinfo{person}{Julie Kientz}, {and} \bibinfo{person}{Alexis Hiniker}.}
  \bibinfo{year}{2018}\natexlab{}.
\newblock \showarticletitle{What makes smartphone use meaningful or
  meaningless?}
\newblock \bibinfo{journal}{\emph{Proceedings of the ACM on Interactive,
  Mobile, Wearable and Ubiquitous Technologies}} \bibinfo{volume}{2},
  \bibinfo{number}{1} (\bibinfo{year}{2018}), \bibinfo{pages}{1--26}.
\newblock


\bibitem[Lyngs et~al\mbox{.}(2022)]%
        {lyngs2022goldilocks}
\bibfield{author}{\bibinfo{person}{Ulrik Lyngs}, \bibinfo{person}{Kai Lukoff},
  \bibinfo{person}{Laura Csuka}, \bibinfo{person}{Petr Slov{\'a}k},
  \bibinfo{person}{Max Van~Kleek}, {and} \bibinfo{person}{Nigel Shadbolt}.}
  \bibinfo{year}{2022}\natexlab{}.
\newblock \showarticletitle{The Goldilocks level of support: Using user
  reviews, ratings, and installation numbers to investigate digital
  self-control tools}.
\newblock \bibinfo{journal}{\emph{International journal of human-computer
  studies}}  \bibinfo{volume}{166} (\bibinfo{year}{2022}),
  \bibinfo{pages}{102869}.
\newblock


\bibitem[Lyngs et~al\mbox{.}(2019)]%
        {lyngs2019self}
\bibfield{author}{\bibinfo{person}{Ulrik Lyngs}, \bibinfo{person}{Kai Lukoff},
  \bibinfo{person}{Petr Slovak}, \bibinfo{person}{Reuben Binns},
  \bibinfo{person}{Adam Slack}, \bibinfo{person}{Michael Inzlicht},
  \bibinfo{person}{Max Van~Kleek}, {and} \bibinfo{person}{Nigel Shadbolt}.}
  \bibinfo{year}{2019}\natexlab{}.
\newblock \showarticletitle{Self-control in cyberspace: Applying dual systems
  theory to a review of digital self-control tools}. In
  \bibinfo{booktitle}{\emph{proceedings of the 2019 CHI conference on human
  factors in computing systems}}. \bibinfo{pages}{1--18}.
\newblock


\bibitem[Lyngs et~al\mbox{.}(2020)]%
        {lyngs2020justhack}
\bibfield{author}{\bibinfo{person}{Ulrik Lyngs}, \bibinfo{person}{Kai Lukoff},
  \bibinfo{person}{Petr Slovak}, \bibinfo{person}{William Seymour},
  \bibinfo{person}{Helena Webb}, \bibinfo{person}{Marina Jirotka},
  \bibinfo{person}{Jun Zhao}, \bibinfo{person}{Max Van~Kleek}, {and}
  \bibinfo{person}{Nigel Shadbolt}.} \bibinfo{year}{2020}\natexlab{}.
\newblock \showarticletitle{'I Just Want to Hack Myself to Not Get Distracted'
  Evaluating Design Interventions for Self-Control on Facebook}. In
  \bibinfo{booktitle}{\emph{Proceedings of the 2020 CHI Conference on Human
  Factors in Computing Systems}}. \bibinfo{pages}{1--15}.
\newblock


\bibitem[Monge~Roffarello and De~Russis(2019)]%
        {monge2019race}
\bibfield{author}{\bibinfo{person}{Alberto Monge~Roffarello} {and}
  \bibinfo{person}{Luigi De~Russis}.} \bibinfo{year}{2019}\natexlab{}.
\newblock \showarticletitle{The race towards digital wellbeing: Issues and
  opportunities}. In \bibinfo{booktitle}{\emph{Proceedings of the 2019 CHI
  conference on human factors in computing systems}}. \bibinfo{pages}{1--14}.
\newblock


\bibitem[Monge~Roffarello and De~Russis(2023a)]%
        {10.1145/3582515.3609523}
\bibfield{author}{\bibinfo{person}{Alberto Monge~Roffarello} {and}
  \bibinfo{person}{Luigi De~Russis}.} \bibinfo{year}{2023}\natexlab{a}.
\newblock \showarticletitle{Nudging Users or Redesigning Interfaces? Evaluating
  Novel Strategies for Digital Wellbeing Through InControl}. In
  \bibinfo{booktitle}{\emph{Proceedings of the 2023 ACM Conference on
  Information Technology for Social Good}} (Lisbon, Portugal)
  \emph{(\bibinfo{series}{GoodIT '23})}. \bibinfo{publisher}{Association for
  Computing Machinery}, \bibinfo{address}{New York, NY, USA},
  \bibinfo{pages}{100–109}.
\newblock
\showISBNx{9798400701160}
\urldef\tempurl%
\url{https://doi.org/10.1145/3582515.3609523}
\showDOI{\tempurl}


\bibitem[Monge~Roffarello and De~Russis(2023b)]%
        {10.1145/3565066.3608703}
\bibfield{author}{\bibinfo{person}{Alberto Monge~Roffarello} {and}
  \bibinfo{person}{Luigi De~Russis}.} \bibinfo{year}{2023}\natexlab{b}.
\newblock \showarticletitle{Nudging Users Towards Conscious Social Media Use}.
  In \bibinfo{booktitle}{\emph{Proceedings of the 25th International Conference
  on Mobile Human-Computer Interaction}} (Athens, Greece)
  \emph{(\bibinfo{series}{MobileHCI '23 Companion})}.
  \bibinfo{publisher}{Association for Computing Machinery},
  \bibinfo{address}{New York, NY, USA}, Article \bibinfo{articleno}{14},
  \bibinfo{numpages}{7}~pages.
\newblock
\showISBNx{9781450399241}
\urldef\tempurl%
\url{https://doi.org/10.1145/3565066.3608703}
\showDOI{\tempurl}


\bibitem[Okeke et~al\mbox{.}(2018b)]%
        {okeke2018goodvibes}
\bibfield{author}{\bibinfo{person}{Fabian Okeke}, \bibinfo{person}{Michael
  Sobolev}, \bibinfo{person}{Nicola Dell}, {and} \bibinfo{person}{Deborah
  Estrin}.} \bibinfo{year}{2018}\natexlab{b}.
\newblock \showarticletitle{Good vibrations: can a digital nudge reduce digital
  overload?}. In \bibinfo{booktitle}{\emph{Proceedings of the 20th
  international conference on human-computer interaction with mobile devices
  and services}}. \bibinfo{pages}{1--12}.
\newblock


\bibitem[Okeke et~al\mbox{.}(2018a)]%
        {okeke2018towards}
\bibfield{author}{\bibinfo{person}{Fabian Okeke}, \bibinfo{person}{Michael
  Sobolev}, {and} \bibinfo{person}{Deborah Estrin}.}
  \bibinfo{year}{2018}\natexlab{a}.
\newblock \showarticletitle{Towards a framework for mobile behavior change
  research}.
\newblock In \bibinfo{booktitle}{\emph{Proceedings of the technology, mind, and
  society}}. \bibinfo{pages}{1--6}.
\newblock


\bibitem[Orzikulova et~al\mbox{.}(2023)]%
        {orzikulova2023finerme}
\bibfield{author}{\bibinfo{person}{Adiba Orzikulova}, \bibinfo{person}{Hyunsung
  Cho}, \bibinfo{person}{Hye-Young Chung}, \bibinfo{person}{Hwajung Hong},
  \bibinfo{person}{Uichin Lee}, {and} \bibinfo{person}{Sung-Ju Lee}.}
  \bibinfo{year}{2023}\natexlab{}.
\newblock \showarticletitle{FinerMe: Examining App-level and Feature-level
  Interventions to Regulate Mobile Social Media Use}.
\newblock \bibinfo{journal}{\emph{Proceedings of the ACM on Human-Computer
  Interaction}} \bibinfo{volume}{7}, \bibinfo{number}{CSCW2}
  (\bibinfo{year}{2023}), \bibinfo{pages}{1--30}.
\newblock


\bibitem[Peper and Harvey(2018)]%
        {peper2018digitaladdiction}
\bibfield{author}{\bibinfo{person}{Erik Peper} {and} \bibinfo{person}{Richard
  Harvey}.} \bibinfo{year}{2018}\natexlab{}.
\newblock \showarticletitle{Digital addiction: Increased loneliness, anxiety,
  and depression}.
\newblock \bibinfo{journal}{\emph{NeuroRegulation}} \bibinfo{volume}{5},
  \bibinfo{number}{1} (\bibinfo{year}{2018}), \bibinfo{pages}{3--3}.
\newblock


\bibitem[Rad and Demeter(2019)]%
        {rad2019youth}
\bibfield{author}{\bibinfo{person}{Dana Rad} {and} \bibinfo{person}{Edgar
  Demeter}.} \bibinfo{year}{2019}\natexlab{}.
\newblock \showarticletitle{Youth Sustainable Digital Wellbeing.}
\newblock \bibinfo{journal}{\emph{Postmodern Openings/Deschideri Postmoderne}}
  \bibinfo{volume}{10}, \bibinfo{number}{4} (\bibinfo{year}{2019}).
\newblock


\bibitem[Roffarello and De~Russis(2023)]%
        {roffarello2023achieving}
\bibfield{author}{\bibinfo{person}{Alberto~Monge Roffarello} {and}
  \bibinfo{person}{Luigi De~Russis}.} \bibinfo{year}{2023}\natexlab{}.
\newblock \showarticletitle{Achieving digital wellbeing through digital
  self-control tools: A systematic review and meta-analysis}.
\newblock \bibinfo{journal}{\emph{ACM Transactions on Computer-Human
  Interaction}} \bibinfo{volume}{30}, \bibinfo{number}{4}
  (\bibinfo{year}{2023}), \bibinfo{pages}{1--66}.
\newblock


\bibitem[Sobolev(2022)]%
        {sobolev2021digitalnudging}
\bibfield{author}{\bibinfo{person}{Michael Sobolev}.}
  \bibinfo{year}{2022}\natexlab{}.
\newblock \showarticletitle{Digital {Nudging}: {Using} {Technology} to {Nudge}
  for {Good}}.
\newblock In \bibinfo{booktitle}{\emph{Behavioral {Science} in the {Wild}}},
  \bibfield{editor}{\bibinfo{person}{Nina Mazar} {and} \bibinfo{person}{Dilip
  Soman}} (Eds.). \bibinfo{publisher}{University of Toronto Press},
  \bibinfo{address}{Toronto, Canada}, \bibinfo{pages}{292--299}.
\newblock


\bibitem[Thomas et~al\mbox{.}(2022)]%
        {thomas2022digital}
\bibfield{author}{\bibinfo{person}{Nisha~M Thomas}, \bibinfo{person}{Sonali~G
  Choudhari}, \bibinfo{person}{Abhay~M Gaidhane},
  \bibinfo{person}{Zahiruddin~Quazi Syed}, \bibinfo{person}{Nisha Thomas},
  {and} \bibinfo{person}{Abhay Gaidhane}.} \bibinfo{year}{2022}\natexlab{}.
\newblock \showarticletitle{‘Digital Wellbeing’: The Need of the Hour in
  Today’s Digitalized and Technology Driven World!}
\newblock \bibinfo{journal}{\emph{Cureus}} \bibinfo{volume}{14},
  \bibinfo{number}{8} (\bibinfo{year}{2022}).
\newblock


\bibitem[Yang et~al\mbox{.}(2019)]%
        {yang2019intention}
\bibfield{author}{\bibinfo{person}{Longqi Yang}, \bibinfo{person}{Michael
  Sobolev}, \bibinfo{person}{Yu Wang}, \bibinfo{person}{Jenny Chen},
  \bibinfo{person}{Drew Dunne}, \bibinfo{person}{Christina Tsangouri},
  \bibinfo{person}{Nicola Dell}, \bibinfo{person}{Mor Naaman}, {and}
  \bibinfo{person}{Deborah Estrin}.} \bibinfo{year}{2019}\natexlab{}.
\newblock \showarticletitle{How intention informed recommendations modulate
  choices: A field study of spoken word content}. In
  \bibinfo{booktitle}{\emph{The World Wide Web Conference}}.
  \bibinfo{pages}{2169--2180}.
\newblock


\bibitem[Zhao et~al\mbox{.}(2022)]%
        {zhao2022avgust}
\bibfield{author}{\bibinfo{person}{Yixue Zhao}, \bibinfo{person}{Saghar
  Talebipour}, \bibinfo{person}{Kesina Baral}, \bibinfo{person}{Hyojae Park},
  \bibinfo{person}{Leon Yee}, \bibinfo{person}{Safwat~Ali Khan},
  \bibinfo{person}{Yuriy Brun}, \bibinfo{person}{Nenad Medvidovi{\'c}}, {and}
  \bibinfo{person}{Kevin Moran}.} \bibinfo{year}{2022}\natexlab{}.
\newblock \showarticletitle{Avgust: automating usage-based test generation from
  videos of app executions}. In \bibinfo{booktitle}{\emph{Proceedings of the
  30th ACM Joint European Software Engineering Conference and Symposium on the
  Foundations of Software Engineering}}. \bibinfo{pages}{421--433}.
\newblock


\bibitem[Zimmermann and Sobolev(2023)]%
        {zimmermann2023digital}
\bibfield{author}{\bibinfo{person}{Laura Zimmermann} {and}
  \bibinfo{person}{Michael Sobolev}.} \bibinfo{year}{2023}\natexlab{}.
\newblock \showarticletitle{Digital Strategies for Screen Time Reduction: A
  Randomized Field Experiment}.
\newblock \bibinfo{journal}{\emph{Cyberpsychology, Behavior, and Social
  Networking}} \bibinfo{volume}{26}, \bibinfo{number}{1}
  (\bibinfo{year}{2023}), \bibinfo{pages}{42--49}.
\newblock


\end{thebibliography}

\end{document}